\documentclass[aps,tightenlines,showpacs]{revtex4}
\usepackage{graphicx}
\usepackage{float}

\newcommand{\bea}{\begin{eqnarray}}
\newcommand{\eea}{\end{eqnarray}}
\newcommand{\vp}{\varphi}
\newcommand{\vG}{{\cal G}}
\newcommand{\xx}{\noindent}

\begin{document}

\title{\Large{The 4PI effective action for $\phi^4$ theory}}

\author{ M.E. Carrington}
\email{carrington@brandonu.ca}
\affiliation{Department of Physics, Brandon University, Brandon, Manitoba, R7A 6A9 Canada\\
and\\
Winnipeg Institute for Theoretical Physics, Winnipeg, Manitoba }

\begin{abstract}
We work with $\phi^4$ theory and study the 4PI effective action at 3-loop order. We discuss the relationship between the equations of motion obtained by taking functional derivatives of the effective action with respect to the variational parameters, and the Schwinger-Dyson (SD) equations. We show that the equation obtained by differentiating with respect to the 2-point function is identical to the SD equation for the 2-point function; differentiating with respect to the 3-point function reproduces the SD equation for the 3-point function, up to 1-loop order; and differentiating with respect to the 4-point function reproduces the SD equation for the 4-point function, at the tree level. These results establish a connection between two techniques for performing non-equilibrium calculations, and provide a starting place for a study of the gauge dependence of quantities derived from an $n$PI effective action.

\end{abstract}


\maketitle

\section{Introduction}

The study of particles in a hot and dense medium is of current interest in the study of plasma physics, condensed matter physics, nuclear physics, and particle physics. Perturbative calculations of observables in hot dense media are often plagued with infrared divergences and other problems. These problems are connected to  the fact that calculations in terms of the bare quantities of the underlying field theory fail to accurately describe the collective behaviour of the hot dense medium, which leads to inconsistencies. One attempts to remedy this situation by working with dressed quantities which take into account the most important collective effects.  An arbitrary resummation procedure however, will produce dressed quantities that violate the conservation laws of the original theory. To avoid this problem, one starts with an action functional that respects the original symmetries. Dressed quantities are obtained from the effective action using a variational procedure which guarentees thermodynamic consistency. Early contributions in this field were made by several authors (see for example \cite{Ward,Lee,Martin,Baym}). The formalism was generalized to relativistic field theories by Cornwall, Jackiw and Tomboulis \cite{CJT}. 

The simplest choice of such an effective action is the 2PI effective action. The full 2PI effective action is completely gauge invarient. In order to do practical calculations however, one must truncate the effective action at some order in  an expansion parameter. This truncated 2PI effective action is also gauge invarient, to the same order in the expansion parameter \cite{ASmit,MEC}, but dressed quantities which are obtained from functional differentiation of a truncated 2PI effective action are not generally gauge invarient. This problem has been discussed by many authors (see for example \cite{Mottola,HvH}). The source of the problem can be understood diagramatically as follows. The 2PI effective theory results in a perturbation theory that uses full resummed propagators and bare vertices. Equivalently, at each order of the perturbation theory, the truncated theory resums certain topologies in preference to others. Since gauge invarience results from the cancellation of gauge dependent terms that occur within different topologies, it is not surprising that the procedure described above produces gauge dependent results. 

Effective theories that involve dressed vertices, as well as dressed propagators, have been proposed \cite{Norton}. An $n$PI effective theory produces a perturbation theory that involves mean fields and dressed 2-point, 3-point $\cdots$ $n$-point functions. A perturbative expansion of this form simultanously resums different classes of topologies. 
As a result, one expects that the problems associated with the gauge dependence of the truncated theory will be reduced: at a given order in the perturbative expansion, gauge dependent contributions should be smaller in a higher order (larger $n$) effective theory.

Problems involving the physics of collective phenomena are also studied within a green's function formalism, in which dressed propagators and vertices follow from a coupled set of Schwinger-Dyson (SD) equations.  As is the case with effective actions, the full hierarchy of SD equations is gauge invarient, but practical calculations involve truncation, and the truncated equations are not gauge invarient. 

It is of interest to understand the connection between these two formalisms. 
In this paper we take a first step in this direction by looking at the 4PI effective action for $\phi^4$ theory. 
We work to 3-loop order and discuss the relationship between the equations of motion obtained by taking functional derivatives of the effective action with respect to the variational parameters, and the SD equations. We show that the equation obtained by differentiation with respect to the 2-point function is identical to the SD equation for the 2-point function. Taking the functional derivative with respect to the 3-point function reproduces the SD equation for the 3-point function, up to 1-loop order. The functional derivative with respect to the 4-point function reproduces the SD equation for the 4-point function, at the tree level. These results establish a connection between two techniques for performing non-equilibrium calculations, and provide a starting place for a study of the gauge dependence of quantities derived from an $n$PI effective action.

This paper is organized as follows. In section II we review the 2PI effective action. We show explicitly that 1PR diagrams cancel at 2-loop order, and that 2PR diagrams cancel at 3-loop order. Of course, these results are well known and are presented only to illustrate the method that we will use in the next section. In section III we study the 4PI effective theory. We work at 3-loop order and obtain an expression for the 4PI effective action as a functional of four variational parameters: the mean field $\vp$, the 2-point function $\vG$, the 3-point function $V$, and the 4-point function $U$. We take functional derivatives of this expression with respect to the variational parameters and discuss the connection between these results and the SD equations. Some conclusions are presented in section IV.

\section{The 2PI effective action}

The generating functional for connected Green functions is given by,
\bea
\label{W-2}
W[J,B] = -i\hbar \ln \int {\cal D}\phi \,{\rm Exp}\,\left[\frac{i}{\hbar}\left(S(\phi)+J_x \phi_x +\frac{1}{2}\phi_x\phi_x
B_{xy}\right)\right]
\eea
Throughout this paper we use DeWitt notation in which it is understood that repeated continuous indices are integrated over:
\bea
J_x \phi_x:=\int d^4 x\, J[x]\,\phi[x] \nonumber
\eea
We expand around the minimum of the exponent which is obtained from the equation
\bea
\label{eom-2}
\frac{\delta S(\phi)}{\delta \phi_x} + J_x + \frac{1}{2} B_{xy} \phi_y \;\Big|_{\phi = \varphi_c} = 0
\eea
We define:
\bea
\label{G0-2}
G^0_{xy} = \big(S_c^{(2)}+B \big)^{-1}_{xy}
\eea
and use the notation 
\bea
S^{(n)}_c = \frac{\delta^n S}{\delta \phi^n}\Big|_{\phi_c}
\eea
We calculate perturbatively by shifting the integration variable in
(\ref{W-2}):
\bea
\phi = \vp_c + \sqrt{\hbar} \chi
\eea
and expanding around the classical trajectory. 

\subsection{2-loop order}

At 2-loops we obtain the familiar result:
\bea
\label{W-2c}
&&W^{(2)}[J,B] = W_0[J,B]+W_1[J,B]+W_2[J,B]\\[2mm]
&& W_0[J,B] = S[\vp_c]+J_x \vp^c_x +\frac{1}{2}\vp^c_x\vp^c_y B_{xy} \nonumber\\[2mm]
&&W_1[J,B] =  i\frac{\hbar}{2}{\rm Tr}\,{\rm ln}\,(G_0^{-1})\nonumber\\[2mm]
&&
\parbox{3.2cm}{$W_2[J,B] = \hbar^2 \Big[-\frac{1}{8} $}
\begin{picture}(45,20)(0,0)
\thicklines
\put(10,0){\circle{20}}\put(30,0){\circle{20}}
\end{picture}
\parbox{.8cm}{$+\frac{1}{12}$}
\begin{picture}(45,20)(0,0)
\thicklines
\put(20,0){\circle{40}}
\put(0,0){\line(40,0){40}}
\end{picture}
\parbox{.8cm}{$+\frac{1}{8}$}
\begin{picture}(62,10)(0,0)
\thicklines
\put(10,0){\circle{20}}
\put(50,0){\circle{20}}
\put(20,0){\line(20,0){20}}
\end{picture}
\parbox{1.5cm}{~~$\Big]_{G_0,S^{(3)}_c,S^{(4)}_c}$} \nonumber
\eea

\vspace*{.5cm}
\xx where the subscript on the square bracket indicates that the lines and vertices in the diagrams represent the propagator  $G_0$ and the vertices $S^{(3)}_c$ and $S^{(4)}_c$. 

\vspace*{1cm}

We want to perform the Legendre transform to obtain the effective action $\Gamma[\vp,{\cal G}]$ where the variables $\vp$ and ${\cal G}$ are defined below.
Taking functional derivatives of (\ref{W-2}) with respect to $J$ and $B$ we obtain,
\bea
\label{phidefns}
&&\frac{\delta W}{\delta J_x} = \langle \phi_x \rangle  =: \varphi_x \\
&&\frac{\delta^2 W}{\delta J_x\delta J_y} = \frac{i}{\hbar}\Big(\langle \phi_x\phi_y\rangle - \varphi_x \varphi_y \Big) \nonumber\\
&&\frac{\delta W}{\delta B_{xy}} = \frac{1}{2}\langle \phi_x\phi_y \rangle \nonumber
\eea
The mean field is defined by the first line in (\ref{phidefns}). To define the 2-point function we start from the definition of the $n$-th Green function:
\bea
\label{Gn}
G^{(n)}(x_1,x_2,\cdots x_n) = \left(-\frac{i}{\hbar}\right)^{n-1}\langle \phi(x_1)\phi(x_2)\cdots \phi(x_n)\rangle
\eea
Subtracting disconnected pieces to obtain the connected 2-point function we have:
\bea
\hbar \vG^{(2)}(x_1,x_2) = -i \big(\langle \phi(x_1)\phi(x_2)\rangle - \varphi(x_1)\vp(x_2)\big)
\eea
and using (\ref{phidefns}) gives:
\bea
\label{Gdefns}
i\hbar \vG^{(2)}_{xy} = -i\hbar \frac{\delta W}{\delta J_x\delta J_y} = 2\frac{\delta W}{\delta B_{xy}}- \vp_x\vp_y
\eea
From now on we drop the superscript $(2)$ on the two point function.
The Legendre transform has the form:
\bea
\label{Gamma-2}
\Gamma[\vp,{\cal G}] = W[J,B]-J_x\vp_x-\frac{1}{2}B_{xy}(i\hbar \vG_{xy}+\vp_x\vp_y)
\eea
This expression is constructed so that the partial functional derivatives with respect to $J$ and $B$ are zero. We obtain:
\bea
\label{Gamma2-ders}
&&\frac{\delta \Gamma}{\delta \vp_x} = -J_x -B_{xy}\vp_y\\
&& \frac{\delta \Gamma}{\delta \vG_{xy}} = -\frac{i\hbar}{2} B_{xy} \nonumber
\eea

We note that $W[J,B]$ is an explicit function of $\vp_c$ and $G_0$ and only dependent on the sources $J$ and $B$ because of the fact that $\vp_c$ and $G_0$ depend on $J$ and $B$, as given in (\ref{eom-2}) and (\ref{G0-2}). We want to remove the dependence on the variables $\vp_c$ and $G_0$ and obtain a function that depends only on $\vp$ and $\vG$. The procedure is as follows. \\

\xx [1]: Calculate a series expansion of the variables $\vp$ and $\vG$ as functions of $\vp_c$ and $G_0$ by using (\ref{phidefns}) and (\ref{Gdefns}) and taking functional derivatives of the perturbative expansion of $W[J,B]$ (Eqn. (\ref{W-2c})).  We obtain series of the form:
\bea
\label{exp1}
&&\vp = \vp^{(0)}[\vp_c,G_0]+\vp^{(1)}[\vp_c,G_0]+\vp^{(2)}[\vp_c,G_0]+\cdots\\
&&{\cal G} = \vG^{(0)}[\vp_c,G_0]+\vG^{(1)}[\vp_c,G_0]+\vG^{(2)}[\vp_c,G_0]+\cdots \nonumber
\eea
To lowest order
\bea
\label{lowestorder}
\vp^{(0)} = \vp_c \,;~~~~\vG^{(0)} = G_0 
\eea
The first order corrections are:
\bea
\label{exppphi}
\parbox{3.5cm}{$\vp^{(1)}[\vp_c,G_0] = -\frac{i\hbar}{2}\Big[$}
\begin{picture}(50,10)(0,0)
\thicklines
\put(20,0){\vector(-1,0){20}}
\put(30,0){\circle{20}}
\end{picture}
\parbox{2cm}{$\Big]_{G_0,S^{(3)}_c,S^{(4)}_c}$}
\eea

\vspace*{.2cm}

\bea
\label{exppG}
\parbox{3.5cm}{$\vG^{(1)}[\vp_c,G_0] = -\frac{i\hbar}{2}\Big[$}
\begin{picture}(50,10)(0,0)
\thicklines
\put(20,0){\vector(-1,0){20}}
\put(20,0){\vector(1,0){20}}
\put(20,10){\circle{20}}
\end{picture}
\parbox{1cm}{$-\frac{1}{2}$}
\begin{picture}(70,10)(0,0)
\thicklines
\put(20,0){\vector(-1,0){20}}
\put(40,0){\vector(1,0){20}}
\put(30,0){\circle{20}}
\end{picture}
\parbox{1cm}{$-\frac{1}{2}$}
\begin{picture}(50,10)(0,0)
\thicklines
\put(20,0){\vector(-1,0){20}}
\put(20,0){\vector(1,0){20}}
\put(20,0){\line(0,1){20}}
\put(20,30){\circle{20}}
\end{picture}
\parbox{2cm}{$\Big]_{G_0,S^{(3)}_c,S^{(4)}_c}$}
\eea
where the arrow on the end of a line indicates explicitly that the line is not truncated. \\

\xx [2]: Invert (\ref{exp1}) by expanding iteratively in powers of $\hbar$ to obtain expressions of the form:
\bea
\label{revexp1}
&&\vp_c = \vp_c^{(0)}[\vp,\vG]+\vp_c^{(1)}[\vp,\vG]+\vp_c^{(2)}[\vp,\vG]+\cdots \\
&&G_0 = G_0^{(0)}[\vp,\vG])+G_0^{(1)}[\vp,\vG]+G_0^{(2)}[\vp,\vG]+\cdots \nonumber
\eea
Substituting (\ref{lowestorder}) and (\ref{exppphi}) into (\ref{exp1}) and keeping terms of order $\hbar$ gives,
\bea
\label{expphi1}
&&\vp_c =  \vp - \vp^{(1)}[\vp,\vG]+{\cal O}(\hbar^2)\\
\label{expG1}
&&G_0 = \vG -   \vG^{(1)}[\vp,\vG]+{\cal O}(\hbar^2)
\eea
and
\bea
\label{expGin1}
(G_0)^{-1}_{xy} = \vG^{-1}_{xy}-\vG^{-1}_{xz}\vG^{(1)}[\vp,\vG]_{zw}\,\vG^{-1}_{wy}+{\cal O}(\hbar^2)
\eea\\
 
 Below we show explicitly that all 1PR diagrams cancel at 2-loop order. Of course, this result is well known. Our intention is only to illustrate the method. We substitute (\ref{W-2c}) into (\ref{Gamma-2}) to obtain:
\bea
\label{Gamma-1c}
&& \Gamma^{(2)}[\vp,\vG] = \underbrace{S[\vp_c]}_\alpha +\underbrace{J_x \vp^c_x}_\beta +\frac{1}{2}\underbrace{\vp^c_x\vp^c_y B_{xy}}_\beta -\underbrace{J_x\vp_x}_\beta-\frac{1}{2}B_{xy}(\underbrace{i\hbar \vG_{xy}}_\gamma+\underbrace{\vp_x\vp_y}_\beta)  + i\frac{\hbar}{2}\underbrace{{\rm Tr}\,{\rm ln}\,(G_0^{-1})}_\gamma \label{Gamma-2b} \\[2mm]
&&
\parbox{1.8cm}{$+\hbar^2 \Big[-\frac{1}{8} $}
\begin{picture}(45,20)(0,0)
\thicklines
\put(10,0){\circle{20}}\put(30,0){\circle{20}}
\end{picture}
\parbox{.8cm}{$+\frac{1}{12}$}
\begin{picture}(45,20)(0,0)
\thicklines
\put(20,0){\circle{40}}
\put(0,0){\line(40,0){40}}
\end{picture}
\parbox{.8cm}{$+\frac{1}{8}$}
\begin{picture}(63,10)(0,0)
\thicklines
\put(10,0){\circle{20}}
\put(50,0){\circle{20}}
\put(20,0){\line(20,0){20}}
\end{picture}
\parbox{2.5cm}{$~~\Big]_{G_0;\,S^{(3)}_c;\,S^{(4)}_c}$}\nonumber 
\eea\\

\vspace*{.2cm}

\xx Note that since we are working to 2-loop order in this section, we can make the replacements $G_0\rightarrow {\cal G}$; $S_c^{(3)}\rightarrow S^{(3)}$ and $S_c^{(4)}\rightarrow S^{(4)}$  inside of all terms in the square bracket.

\vspace*{1cm}

We group the terms as indicated, and consider each set separately. \\

\xx [$\alpha$]. The term marked $\alpha$ can be expanded:
\bea
\label{alpha1}
(\alpha) = S[\vp_c] = S[\vp+(\vp_c-\vp)] = S[\vp]+(\vp_c-\vp)_x\frac{\delta S}{\delta \vp_x} +\frac{1}{2}(\vp_c-\vp)_x(\vp_c-\vp)_y\frac{\delta^2 S}{\delta \vp_x\delta \vp_y}
\eea

\xx [$\beta$]. The terms marked $\beta$ give:
\bea
\label{beta1}
&&(\beta) = (J_x+B_{xy}\vp^c_y)(\vp_c-\vp)_x-\frac{1}{2}B_{xy}(\vp_c-\vp)_x(\vp_c-\vp)_y \\[2mm]
&&~~~~ = -\frac{\delta S}{\delta \vp_x^c}(\vp_c-\vp)_x-\frac{1}{2}B_{xy}(\vp_c-\vp)_x(\vp_c-\vp)_y\nonumber\\[2mm]
&&~~~~=-\left(\frac{\delta S}{\delta \vp_x}+(\vp_c-\vp)_y\frac{\delta^2 S}{\delta \vp_x\,\delta \vp_y}\right)(\vp_c-\vp)_x-\frac{1}{2}B_{xy}(\vp_c-\vp)_x(\vp_c-\vp)_y\nonumber
\eea
where we have used (\ref{eom-2}) to go from the first to the second line and made the expansion 
\bea
\label{tadex}
\frac{\delta S}{\delta \vp_x^c} = \frac{\delta S}{\delta \vp_x}+(\vp_c-\vp)_y\frac{\delta S}{\delta \vp_x\,\delta \vp_y}
\eea
in going from the second to third lines. 
The second term in (\ref{alpha1}) and the first term in (\ref{beta1}) cancel exactly. Note that it is because of this cancellation that we do not need the term ${\cal O}(\hbar^2)$ which is dropped in (\ref{expphi1}). \\

\xx [$\gamma$]. We use (\ref{G0-2}) to eliminate the source $B$. The factors $G_0^{-1}$ are expanded using (\ref{expGin1}) and $S^{(2)}_c$ is expanded as above. We obtain,
\bea
\label{gamma1}
(\gamma) = i\frac{\hbar}{2}{\rm Tr ~ ln}\;\vG^{-1} - i\frac{\hbar}{2}\,\vG_{xy}\,\Big(\vG_{xy}^{-1}-\Big[\frac{\delta^2 S}{\delta \vp_x\,\delta \vp_y}+(\vp_c-\vp)_z \frac{\delta^3 S}{\delta \vp_x\,\delta \vp_y\delta\vp_z}\Big]\Big)
\eea
Contributions to 1PR diagrams come from the last term in (\ref{alpha1}), the last two terms in (\ref{beta1}), and the last term in (\ref{gamma1}). Combining we have:
\bea
-\frac{1}{2}\underbrace{(\vp_c-\vp)_x}_{-\vp^{(1)}[\vp,\vG]}\underbrace{(\vp_c-\vp)_y}_{-\vp^{(1)[\vp,\vG]}}\left(\underbrace{\frac{\delta^2 S}{\delta \vp_x\,\delta\vp_y}+B_{xy}}_{\vG}\right)+i\frac{\hbar}{2}\vG_{xy} \underbrace{(\vp_c-\vp)_z}_{-\vp^{(1)}[\vp,\vG]} \frac{\delta^3 S}{\delta \vp_x\,\delta \vp_y\,\delta \vp_z}
\eea
where the substitutions indicated by the underbraces are valid up to corrections of order $\hbar^3$ (using (\ref{G0-2}), (\ref{lowestorder}), (\ref{exppphi}) and (\ref{expphi1})). Using (\ref{exppphi}) it is easy to see that both terms correspond to the dumbell graph. The first carries a factor of $\hbar^2/8$ and the second a factor of $-\hbar^2/4$. Since symmetry factor for the dumbell graph in (\ref{Gamma-1c}) is 1/8, we find that, after summing all contributions, the dumbell graphs cancel. This is merely the well known statement that the 2PI effective action is one-particle-irreducible.  Since there are no 2PR diagrams at 2-loop order we have to go to next order in the loop expansion (order $\hbar^3$) in order to see cancellations of 2PR graphs. This is the subject of the next section.\\

\subsection{3-loop order}

In this section we will calculate the 2PI effective action to  3-loop order and explicitly demonstrate the cancellation of the 2PR diagrams. Of course, it is well known that both the 1PR and 2PR diagrams cancel in the 2PI effective action \cite{CJT}. We remind the reader that our purpose is to demonstrate the technique that will be used in the next section to study the 4PI effective action. Throughout this section we will ignore all 1PR diagrams. Equivalently, we assume that we are allowed to make the substitution $\vp_c\rightarrow\vp$ everywhere.  Our goal is to demonstrate the cancellation of the 2PR diagrams, without bothering with the 1PR diagrams which can be shown to cancel using the same method as in the previous section. 

In order to simplify the equations, we introduce some shorthand notation. We define:
\bea
\frac{\delta^2 S}{\delta \vp_x\delta \vp_y} = S^{(2)}_{xy}\,;~~~\frac{\delta^3 S}{\delta \vp_x\delta \vp_y\delta\vp_z} = S^{(3)}_{xyz}\,;~~~{\rm etc.}
\eea
In addition, whenever possible to do so without causing confusion, we will use a type of matrix notation. We give some examples below:
\bea
\label{short}
&& J_x \vp_x \rightarrow J\vp \\[2mm]
&& (\vp_c-\vp)_x(\vp_c-\vp)_y B_{xy} \rightarrow (\vp_c-\vp)^2 B\nonumber
\eea\\
Note that in principle an expression like $(\vp_c-\vp)^2 B$ is ambiguous since it could mean either 
\bea
(\vp_c-\vp)_x(\vp_c-\vp)_y B_{xy}~~{\rm or}~~ (\vp_c-\vp)_x(\vp_c-\vp)_x B_{yy}
\eea
Since we are only dealing with connected diagrams it is clear that the first expression is to be used. The only place where this shorthand notation can cause confusion is when a given combination of factors  could correspond to either of two diagrams with different topologies. In these cases, we will  write  the indices explicitly. \\

\xx In order to obtain an expression for $\Gamma^{(3)}[\vp,\vG]$ we use (\ref{Gamma-2}) and expand $W[J,B]$ to three loop order: $W[J,B] = W_0[J,B]+W_1[J,B]+W_2[J,B]+W_3[J,B]$. We group terms as in (\ref{Gamma-2b}) and use expressions of the form (\ref{revexp1}) to remove the dependence on the variables $\vp_c$ and $G_0$ and obtain a function that depends only on $\vp$ and $\vG$.\\

\xx There are three different places where 3-loop graphs will appear:\\

\xx [1] Contributions from $W_3[J,B]$.\\

\xx [2] When working at order $\hbar^3$ it is not sufficient to replace $G_0$ with $\vG$ in the contributions from $W_2[J,B]$, as in the previous section. Instead we must use (\ref{expG1}) and write $G_0 = \vG -   \vG^{(1)}[\vp,\vG]+{\cal O}(\hbar^2)$. Note that in principle the vertices should also be expanded by writing $S_c^{(n)} = S^{(n)}[\vp +(\vp_c-\vp)]$ and Taylor expanding as in (\ref{tadex}). However, since we are ignoring 1PR diagrams, we make the replacement $S^{(n)}_c\rightarrow S^{(n)}$. \\

\xx [3] In the terms marked $(\gamma)$, we use (\ref{G0-2}) to remove the source $B$, make the replacement $\vp_c\rightarrow\vp$, and expand $G_0$ to second order.  As will be seen below, this expansion involves terms of the form $(\vG^{(1)})^2$ but terms proportional to $\vG^{(2)}$ cancel, and thus we will not need anything more than $\vG^{(1)}$ as given in (\ref{exppG}) and (\ref{expG1}).\\

\xx We calculate below each of these three contributions.\\

\xx [1] The contribution from $W_3[J,B]$ (neglecting 1PR graphs) is:
\bea
\label{W3}
\parbox{3.2cm}{$W_3[J,B] = i\hbar^3\Big[\frac{1}{16}$}
\begin{picture}(70,20)(0,0)
\thicklines
\put(10,0){\circle{20}}
\put(30,0){\circle{20}}
\put(50,0){\circle{20}}
\end{picture}
\parbox{1cm}{$-\frac{1}{8}$}
\begin{picture}(60,20)(0,0)
\thicklines
\put(10,0){\circle{20}}
\put(30,0){\circle{20}}
\put(30,-10){\line(0,1){20}}
\end{picture}
\parbox{1cm}{$+\frac{1}{16}$}
\begin{picture}(60,20)(0,0)
\thicklines
\put(20,0){\circle{40}}
\put(4.7,10.2){\line(1,0){30.2}}
\put(4.7,-10.2){\line(1,0){30.2}}
\end{picture}\\[8mm]
\parbox{1cm}{$+\frac{1}{24}$}
\begin{picture}(60,20)(0,0)
\thicklines
\put(20,0){\circle{40}}
\put(0,0){\line(1,0){40}}
\put(20,0){\line(0,-1){20}}
\end{picture}
\parbox{1cm}{$-\frac{1}{8}$}
\begin{picture}(60,20)(0,0)
\thicklines
\put(20,0){\circle{40}}
\put(10,0){\circle{20}}
\put(20,0){\line(1,0){20}}
\end{picture}
\parbox{1cm}{$+\frac{1}{48}$}
\begin{picture}(75,20)(0,0)
\thicklines
\put(20,0){\circle{40}}
\put(45,0){\circle{40}}
\end{picture}
\parbox{2cm}{$\Big]_{G_0;\,S^{(3)}_c;\,S^{(4)}_c}$}\nonumber
\eea

\vspace*{1cm}

\xx Note that the first three diagrams are 2PR and the second three are 2PI. Since we are working to 3-loop order in this section, we can make the replacements $G_0\rightarrow {\cal G}$; $S_c^{(3)}\rightarrow S^{(3)}$ and $S_c^{(4)}\rightarrow S^{(4)}$ inside of all terms in the square bracket.  \\

\xx [2] Consider the graphs that we obtain by substituting (\ref{exppG}) and (\ref{expG1}) into $W_2[J,B]$ (ignoring 1PR terms). The result can be represented graphically:
\bea
\label{expW2}
\parbox{1.2cm}{$i\frac{\hbar^3}{8}\Big[2$}
\begin{picture}(60,20)(0,0)
\thicklines
\put(10,0){\circle{20}}
\put(30,0){\circle{20}}
\put(30,-10){\line(0,1){20}}
\end{picture}
\parbox{1cm}{$-$}
\begin{picture}(60,20)(0,0)
\thicklines
\put(20,0){\circle{40}}
\put(4.7,10.2){\line(1,0){30.2}}
\put(4.7,-10.2){\line(1,0){30.2}}
\end{picture}
\parbox{1cm}{$-$}
\begin{picture}(70,20)(0,0)
\thicklines
\put(10,0){\circle{20}}
\put(30,0){\circle{20}}
\put(50,0){\circle{20}}
\end{picture}
\parbox{1cm}{$\Big]_{\vG;\,S^{(3)};\,S^{(4)}}$}
\eea\\

\xx [3] Finally, we consider the expansion of the terms marked $(\gamma)$ in (\ref{Gamma-2b}).  We use (\ref{G0-2}) to eliminate the source $B$ and make the replacement $\vp_c\rightarrow \vp$ since doing so means dropping 1PR graphs. We want to expand the propagator using an expression like (\ref{expGin1}), but working to one higher power in $\hbar$. \\

\xx  Eliminating $B$ we have:
\bea
(\gamma) = \frac{i\hbar}{2}\Big(\vG S^{(2)} + \big[ {\rm Tr\,ln}\,G_0^{-1}-\vG G_0^{-1}\Big]\Big)
\eea
Expanding the term in square brackets gives
\bea
[~~]\rightarrow {\rm Tr\,ln}\,\vG^{-1}-\vG_{xy}\vG^{-1}_{yx}-\frac{1}{2}\big(\vG^{(1)}_{xy} \vG^{-1}_{yz}\vG^{(1)}_{zw}\vG^{-1}_{wx}\big)
\eea
where we have used 
\bea
\label{expGin2}
(G_0)^{-1}_{xy} = \vG^{-1}_{xy}-\vG^{-1}_{xz}\vG^{(1)}_{zw}\,\vG^{-1}_{wy} - \vG^{-1}_{xz}\vG^{(2)}_{zw}\vG^{-1}_{wy} 
\eea
Note that the terms containing the factor $\vG^{(2)}$ cancel exactly.
Substituting we obtain, 
\bea
(\gamma) = i\frac{\hbar}{2}\Big(\vG \big[S^{(2)}-\vG^{-1}\big]+{\rm Tr\,ln\,}\vG^{-1}-\frac{1}{2}\big(\vG^{(1)}_{xy} \vG^{-1}_{yz}\vG^{(1)}_{zw}\vG^{-1}_{wx}\big)\Big)
\eea
Consider the last term in this expression. This term is already explicitly of order $\hbar^3$ and therefore we can make the replacements $\vp_c\rightarrow\vp$ and $G_0\rightarrow\vG$ in the arguments of $\vG^{(1)}$. 
Using the 1PI part of (\ref{exppG}) we obtain a result that can be represented graphically:
\bea
\label{expgamma}
\parbox{1.8cm}{$i\frac{\hbar^3}{16}\Big[-2$}
\begin{picture}(60,20)(0,0)
\thicklines
\put(10,0){\circle{20}}
\put(30,0){\circle{20}}
\put(30,-10){\line(0,1){20}}
\end{picture}
\parbox{1cm}{$+$}
\begin{picture}(60,20)(0,0)
\thicklines
\put(20,0){\circle{40}}
\put(4.7,10.2){\line(1,0){30.2}}
\put(4.7,-10.2){\line(1,0){30.2}}
\end{picture}
\parbox{1cm}{$+$}
\begin{picture}(70,20)(0,0)
\thicklines
\put(10,0){\circle{20}}
\put(30,0){\circle{20}}
\put(50,0){\circle{20}}
\end{picture}
\parbox{1cm}{$\Big]_{\vG;\,S^{(3)};\,S^{(4)}}$}
\eea\\

\xx Combining (\ref{W3}), (\ref{expW2}) and (\ref{expgamma}) we see that the 2PR terms cancel. We are left with:
\bea
&&\parbox{12.6cm}{$\Gamma^{(3)}[\vp,\vG] = S[\vp]+i\frac{\hbar}{2}\Big(\vG \big[S^{(2)}-\vG^{-1}\big]+{\rm Tr\,ln\,}\vG^{-1}\Big)$} \\[4mm]
&&~~~~\parbox{1.8cm}{$+\hbar^2 \Big[-\frac{1}{8} $}
\begin{picture}(45,20)(0,0)
\thicklines
\put(10,0){\circle{20}}\put(30,0){\circle{20}}
\end{picture}
\parbox{.8cm}{$+\frac{1}{12}$}
\begin{picture}(45,20)(0,0)
\thicklines
\put(20,0){\circle{40}}
\put(0,0){\line(40,0){40}}
\end{picture}
\parbox{1.5cm}{~~$\Big]_{\vG,S^{(3)},S^{(4)}}$} \nonumber\\[8mm]
&&~~~~\parbox{1.7cm}{$i\hbar^3\Big[~\frac{1}{24}$}
\begin{picture}(60,20)(0,0)
\thicklines
\put(20,0){\circle{40}}
\put(0,0){\line(1,0){40}}
\put(20,0){\line(0,-1){20}}
\end{picture}
\parbox{1cm}{$-\frac{1}{8}$}
\begin{picture}(60,20)(0,0)
\thicklines
\put(20,0){\circle{40}}
\put(10,0){\circle{20}}
\put(20,0){\line(1,0){20}}
\end{picture}
\parbox{1cm}{$+\frac{1}{48}$}
\begin{picture}(75,20)(0,0)
\thicklines
\put(20,0){\circle{40}}
\put(45,0){\circle{40}}
\end{picture}
\parbox{2cm}{$\Big]_{\vG;\,S^{(3)};\,S^{(4)}}$}\nonumber\\[6mm]\nonumber
\eea

\section{4PI Effective Theory}

\subsection{Generating Functional}

The generating functional for connected diagrams is given by:
\bea
\label{W-4}
W[J,B,K,H] = -i\hbar\,{\rm ln}\,\int {\cal D}\phi\;{\rm Exp}\,\Big[\frac{i}{\hbar}\Big(S(\phi)+J_x\phi_x+\frac{1}{2}B_{xy}\phi_x\phi_y+\frac{1}{3!}K_{xyz}\phi_x\phi_y\phi_z+\frac{1}{4!}H_{xyzw}\phi_x\phi_y\phi_z\phi_w\Big)\Big]
\eea
where $K_{xyz}$ and $H_{xyzw}$ are sources. 
We expand around the solutions to the classical equations of motion by shifting
\bea
\phi = \vp_c+\sqrt{\hbar}\chi
\eea
where $\phi_c$ is defined by the equation:
\bea
\frac{\delta S}{\delta \phi}\Big|_{\vp_c} + J + B \vp_c +\frac{1}{2}K \vp_c^2+\frac{1}{3!}H \vp_c^3 = 0
\eea
In this section, many of the variables, like $\vp_c$ as defined above, have counterparts for the 2PI effective action discussed in the previous section. To avoid a proliferation of indices we will use the same variable, and not define a new variable with an extra index (like $\vp^c_{4PI}$). We assume always that variables used in this section correspond to the definitions made within this section (unless otherwise indicated).
Shifting around this classical solution it is easy to show that the perturbative expansion of $W[J,B,K,H]$ involves a propagator and vertices given by: 
\bea
\label{prop-4}
&&G_0^{-1}[\vp_c] = S_c^{(2)}+B+K\vp_c+\frac{1}{2}H\vp_c^2 \\
&& \tilde V_0[\vp_c] = S^{(3)}_c+K+H\vp_c \nonumber \\[2mm]
&& \tilde U_0[\vp_c] = S^{(4)}_c+H \nonumber
\eea
where the tilde's indicate truncated vertices. \\

Taking functional derivatives of $W[J,B,K,H]$ we obtain,
\bea
&&\vp:=\langle \phi\rangle = \frac{\delta W}{\delta J} \\
&&i\hbar \vG := -i\hbar \frac{\delta^2 W}{\delta J^2}  = 2\frac{\delta W}{\delta B} - \vp^2 \nonumber \\
&& -\hbar^2 V := -\hbar^2 \frac{\delta^3 W}{\delta J^3} = 3!\frac{\delta W}{\delta K}-3i\hbar \vG \vp -\vp^3 \nonumber \\
&& -i\hbar^3 U := i\hbar^3 \frac{\delta^4 W}{\delta J^4} = 4!\frac{\delta W}{\delta H}+4\hbar^2 V \vp +3\hbar^2 \vG^2-6i\hbar \vG \vp^2 -\vp^4 \nonumber
\eea
where $\vp$, $\vG$, $V$, and $U$ are the connected 1-point function, 2-point function, 3-point function and 4-point function, respectively, and we have used ({\ref{Gn}).

\subsection{The Effective Action}

The effective action is obtained by Legendre transforming the generating functional. The Legendre transformation is constructed so that partial derivatives with respect to the sources $J$, $B$, $K$ and $H$ are zero. We make the definition:
\bea
\label{Gamma-4}
&&\Gamma[\vp,\vG,V,U] \\[2mm]
&&~~~~= W[J,B,K,H] \nonumber \\[2mm]
&&~~~~-J\vp-\frac{1}{2} B(i\hbar \vG+\vp^2)-\frac{1}{3!}K(-\hbar^2 V+3i\hbar \vG \vp + \vp^3) -\frac{1}{4!}H(-i\hbar^3 U -4\hbar^2 V \vp -3 \hbar^2 \vG^2 +6i \hbar \vG \vp^2 +\vp^4)\nonumber
\eea
Taking partial derivatives we obtain the equations of motion:
\bea
&&\frac{\delta \Gamma}{\delta \vp} = -J -B\vp-\frac{1}{2}K\vp^2-\frac{1}{3!}H\vp^3 \\
&&\frac{\delta \Gamma}{\delta \vG} = -\frac{i\hbar}{2}\left(B+K\vp+H\vp^2\right)+\frac{\hbar^2}{3}H\vG \nonumber \\
&&\frac{\delta \Gamma}{\delta V} = \frac{\hbar^2}{3!}\left(K+H\vp\right)\nonumber \\
&&\frac{\delta \Gamma}{\delta U} = i\hbar^3 H\nonumber
\eea

As in the previous section, the goal is to obtain perturbative expansions for the variables $\{\vp_c,G_0,\tilde V_0,\tilde U_0\}$ in terms of the variables $\{\vp,\vG,V,U\}$. We proceed as before. The first step is to derive expressions analogous to (\ref{exp1}) by taking functional derivatives of the perturbative expression for $W[J,B,K,H]$. This procedure leads to expansions of the form:
\bea
\label{exp2}
&&\vp = \vp^{(0)}[\vp_c,G_0,\tilde V_0,\tilde U_0]+\vp^{(1)}[\vp_c,G_0,\tilde V_0,\tilde U_0]+\vp^{(2)}[\vp_c,G_0,\tilde V_0,\tilde U_0]+\cdots\\
&&{\cal G} = \vG^{(0)}[\vp_c,G_0,\tilde V_0,\tilde U_0]+\vG^{(1)}[\vp_c,G_0,\tilde V_0,\tilde U_0]+\vG^{(2)}[\vp_c,G_0,\tilde V_0,\tilde U_0]+\cdots \nonumber\\
&&{V} = V^{(0)}[\vp_c,G_0,\tilde V_0,\tilde U_0]+V^{(1)}[\vp_c,G_0,\tilde V_0,\tilde U_0]+V^{(2)}[\vp_c,G_0,\tilde V_0,\tilde U_0]+\cdots \nonumber\\
&&U = U^{(0)}[\vp_c,G_0,\tilde V_0,\tilde U_0]+U^{(1)}[\vp_c,G_0,\tilde V_0,\tilde U_0]+U^{(2)}[\vp_c,G_0,\tilde V_0,\tilde U_0]+\cdots \nonumber
\eea
 The second step is to invert (\ref{exp2}) by expanding iteratively in powers of $\hbar$ and obtain the equations of the form:
\bea
\label{revexp2}
&&\vp_c = \vp_c^{(0)}[\vp,\vG,V,U]+\vp_c^{(1)}[\vp,\vG,V,U]+\vp_c^{(2)}[\vp,\vG,V,U]+\cdots \\
&&G_0 = G_0^{(0)}[\vp,\vG,V,U])+G_0^{(1)}[\vp,\vG,V,U]+G_0^{(2)}[\vp,\vG,V,U]+\cdots \nonumber\\
&&\tilde V_0 = \tilde V_0^{(0)}[\vp,\vG,V,U])+\tilde V^{(1)}[\vp,\vG,V,U]+\tilde V^{(2)}[\vp,\vG,V,U]+\cdots \nonumber\\
&&\tilde U_0 = \tilde U_0^{(0)}[\vp,\vG,V,U])+\tilde U_0^{(1)}[\vp,\vG,V,U]+\tilde U_0^{(2)}[\vp,\vG,V,U]+\cdots \nonumber
\eea

The cancellation of the 1PR and 2PR diagrams works essentially the same as before. The interesting question is what happens to the 3PR and 4PR diagrams. To study this point, we will work at order $\hbar^3$ and assume that all 1PR and 2PR diagrams cancel. Effectively, we make the replacements $\vp_c\rightarrow \vp$ and $G_0\rightarrow \vG$ everywhere.  The 4PI effective action is given by (\ref{W-4}) and (\ref{Gamma-4}).  Dropping 1PR and 2PR diagrams we obtain (to order $\hbar^3$): 
\bea
\label{Gamma-4a}
&&\Gamma^{(3)}[\vp,\vG,V,U] = \underbrace{S[\vp_c]}_{\alpha}+\underbrace{J \vp_c}_{\beta} +\frac{1}{2}\underbrace{\vp_c\vp_c B}_{\beta} +\frac{1}{3!}\underbrace{K\vp_c^3}_{\beta}+\frac{1}{4!}\underbrace{H\vp_c^4}_{\beta} + i\frac{\hbar}{2}\underbrace{{\rm Tr}\,{\rm ln}\,(G_0^{-1})}_{\gamma}\nonumber\\[2mm]
&&~~~~~~
\parbox{1.8cm}{$+\hbar^2 \Big[-\frac{1}{8} $}
\begin{picture}(45,20)(0,0)
\thicklines
\put(10,0){\circle{20}}\put(30,0){\circle{20}}
\end{picture}
\parbox{.8cm}{$+\frac{1}{12}$}
\begin{picture}(45,20)(0,0)
\thicklines
\put(20,0){\circle{40}}
\put(0,0){\line(40,0){40}}
\end{picture}
\parbox{1.5cm}{~~$\Big]_{G_0;\,S_c^{(3)};\,S_c^{(4)}}$} \nonumber\\[3mm]
\\
&&~~~~\parbox{2cm}{$i\hbar^3\Big[\frac{1}{24}$}
\begin{picture}(60,20)(0,0)
\thicklines
\put(20,0){\circle{40}}
\put(0,0){\line(1,0){40}}
\put(20,0){\line(0,-1){20}}
\end{picture}
\parbox{1cm}{$-\frac{1}{8}$}
\begin{picture}(60,20)(0,0)
\thicklines
\put(20,0){\circle{40}}
\put(10,0){\circle{20}}
\put(20,0){\line(1,0){20}}
\end{picture}
\parbox{1cm}{$+\frac{1}{48}$}
\begin{picture}(75,20)(0,0)
\thicklines
\put(20,0){\circle{40}}
\put(45,0){\circle{40}}
\end{picture}
\parbox{2cm}{$\Big]_{G_0;\,S_c^{(3)};\,S_c^{(4)}}$}\nonumber\\[10mm]\nonumber
&&~~~~-\underbrace{J\vp}_{\beta}-\frac{1}{2} B(\underbrace{i\hbar \vG}_{\gamma}+\underbrace{\vp^2}_{\beta})-\frac{1}{3!}K(\underbrace{-\hbar^2 V}_{\epsilon}+\underbrace{3i\hbar \vG \vp}_{\gamma} + \underbrace{\vp^3}_{\beta}) \nonumber\\[2mm]
&&~~~~-\frac{1}{4!}H(\underbrace{-i\hbar^3 U}_{\tau} -\underbrace{4\hbar^2 V \vp}_{\epsilon} -\underbrace{3 \hbar^2 \vG^2}_{\tau} +\underbrace{6i \hbar \vG \vp^2}_{\gamma} +\underbrace{\vp^4}_{\beta})\nonumber
\eea
We start by considering the terms marked $(\alpha)$, $(\beta)$ and $(\gamma)$. Using (\ref{prop-4}) and making the replacements $\vp_c\rightarrow \vp$ and $G_0\rightarrow \vG$ we have:
\bea
&&(\alpha) = S[\vp]\\[2mm]
&&(\beta) = 0\nonumber\\[2mm]
&&(\gamma) = i\frac{\hbar}{2}\Big({\rm Tr\,ln\,}\vG^{-1}+\vG\big[S^{(2)}-\vG^{-1}\big]\Big)\nonumber
\eea
Combining these results gives:
\bea
\label{part1}
\Gamma[\vp,\vG,V,U]_{{\rm Part}\,1} = S[\vp]+i\frac{\hbar}{2}\Big({\rm Tr\,ln\,}\vG^{-1}+\vG\big[S^{(2)}-\vG^{-1}\big]\Big)
\eea

\xx The interesting contributions come from: \\

\xx [1] The 3-loop diagrams in (\ref{Gamma-4a}). \\

\xx [2] Terms obtained from the 2-loop diagrams in (\ref{Gamma-4a}) with the vertices expanded to order $\hbar$.\\

\xx [3] Contributions to the terms in (\ref{Gamma-4a}) marked $\epsilon$ and $\tau$ with the sources removed using (\ref{prop-4}) and the vertices expanded to order $\hbar$.\\

\xx Using (\ref{prop-4}) to remove the sources we have:
\bea
\label{Gamma-4b}
\Gamma[\vp,\vG,V,U]_{{\rm Part}\,2} = &&\frac{\hbar^2}{3!} V(\underbrace{\tilde V_0}_b-\underbrace{S^{(3)}_c}_{e})\\[2mm]
&& i\frac{\hbar^3}{4!}U(\underbrace{\tilde U_0}_c - \underbrace{S^{(4)}_c}_{e})+\frac{\hbar^2}{8}\vG^2 (\underbrace{\tilde U_0}_a-\underbrace{S^{(4)}_c}_{e}) \nonumber\\[10mm]
&&
\parbox{1.8cm}{$+\hbar^2 \Big[-\frac{1}{8} $}
\begin{picture}(45,20)(0,0)
\thicklines
\put(10,0){\circle{20}}\put(30,0){\circle{20}}
\end{picture}
\parbox{.8cm}{$+\frac{1}{12}$}
\begin{picture}(45,20)(0,0)
\thicklines
\put(20,0){\circle{40}}
\put(0,0){\line(40,0){40}}
\end{picture}
\parbox{1.5cm}{~~$\Big]_{G_0,\tilde V_0,\tilde U_0}$} \nonumber\\[0mm]\nonumber\\
&& ~~~~~~~~~~~~~\underbrace{~~~~~~~~~~~~~~}_a~+~\underbrace{~~~~~~~~~~~~~~}_b \nonumber \\
&&\parbox{1.8cm}{$+ i\hbar^3\Big[\frac{1}{24}$}
\begin{picture}(55,20)(0,0)
\thicklines
\put(20,0){\circle{40}}
\put(0,0){\line(1,0){40}}
\put(20,0){\line(0,-1){20}}
\end{picture}
\parbox{1cm}{$-\frac{1}{8}$}
\begin{picture}(55,20)(0,0)
\thicklines
\put(20,0){\circle{40}}
\put(10,0){\circle{20}}
\put(20,0){\line(1,0){20}}
\end{picture}
\parbox{1cm}{$+\frac{1}{48}$}
\begin{picture}(65,20)(0,0)
\thicklines
\put(20,0){\circle{40}}
\put(45,0){\circle{40}}
\end{picture}
\parbox{2cm}{$~~~~\Big]_{G_0,\tilde V_0,\tilde U_0}$}\nonumber\\[8mm]
&& ~~~~~~~~~~~~~\underbrace{~~~~~~~~~~~~~~}_c~~~~+~~~~\underbrace{~~~~~~~~~~~~~~}_c \nonumber ~~~~+~~~~\underbrace{~~~~~~~~~~~~~~~~}_c \nonumber
\eea\\
Note that the terms marked (a) cancel (up to 2PR terms).\\

\xx The terms marked (e) are already functions of $\vG$, $V$ and $U$.\\

\xx We need to rewrite the remaining terms as functions of $\{\vp,\vG,V,U\}$ instead of $\{\vp_c,G_0,\tilde V_0,\tilde U_0\}$. We take functional derivatives of the perturbative expansion of $W[J,B,K,H]$ and find equations of the form (\ref{exp2}), which give expansions of $V$ and $U$ in terms of $\{\vp_c,G_0,\tilde V_0,\tilde U_0\}$.  Recall that since we are ignoring 1PR and 2PR terms we can make the replacements $\vp_c\rightarrow \vp$ and $G_0\rightarrow \vG$. We obtain:
\bea
\label{expV}
&&\parbox{2cm}{$V~~ = \Big[~~-$}
\begin{picture}(50,20)(0,0)
\thicklines
\put(20,0){\vector(1,0){20}}
\put(20,0){\vector(-1,1){15}}
\put(20,0){\vector(-1,-1){15}}
\end{picture}
\parbox{1.5cm}{$-i\hbar\;\Big($}
\begin{picture}(70,20)(0,0)
\thicklines
\put(20,0){\circle{40}}
\put(40,0){\vector(1,0){20}}
\put(2.5,-8){\vector(-1,-1){15}}
\put(2.5,8){\vector(-1,1){15}}
\end{picture}
\parbox{1.6cm}{$-\frac{3}{2}$}
\begin{picture}(70,20)(0,0)
\thicklines
\put(20,0){\circle{40}}
\put(40,0){\vector(1,0){20}}
\put(0,0){\vector(-1,-1){15}}
\put(0,0){\vector(-1,1){15}}
\end{picture} 
\parbox{1.9cm}{$\Big)\Big]_{G_0,\tilde V_0,\tilde U_0}$}\\[8mm]
\label{expU}
&&\parbox{3cm}{$U ~~= ~~\Big[~~-~~~~~~~$}
\begin{picture}(40,20)(0,0)
\thicklines
\put(0,0){\vector(1,1){15}}
\put(0,0){\vector(1,-1){15}}
\put(0,0){\vector(-1,1){15}}
\put(0,0){\vector(-1,-1){15}}
\end{picture}
\parbox{1.5cm}{$+3$}
\begin{picture}(70,20)(0,0)
\thicklines
\put(0,0){\line(1,0){20}}
\put(20,0){\vector(1,1){15}}
\put(20,0){\vector(1,-1){15}}
\put(0,0){\vector(-1,-1){15}}
\put(0,0){\vector(-1,1){15}}
\end{picture}
\parbox{2cm}{$\Big]_{G_0,\tilde V_0,\tilde U_0}$}\\[6mm]\nonumber
\eea
Note that in the equations above the two factors of  `3' both come from the fact that there are three different permutations of the external legs for the corresponding diagrams. We must be careful when using these vertex corrections to produce diagrams, since different permutations of external legs will produce diagrams with different topologies. \\

\xx We define truncated vertices as follows:
\bea
\label{trun-vert}
\bar V = \vG^{-3}V\,,~~ \bar U = \vG^{-4} U
\eea
In order to rewrite the terms marked (b) we invert (\ref{expV}) to obtain:
\bea
\label{Vbar1}
\tilde V_0 = -\bar V+\vG^{-3} V^{(1)} = -\bar V+\bar V^{(1)}
\eea
with
\bea
\label{V1}
\parbox{2.8cm}{$\bar V^{(1)} = -i\hbar \Big[$}
\begin{picture}(70,20)(0,0)
\thicklines
\put(20,0){\circle{40}}
\put(40,0){\line(1,0){20}}
\put(2.5,-8){\line(-1,-1){15}}
\put(2.5,8){\line(-1,1){15}}
\end{picture}
\parbox{1.6cm}{$-\frac{3}{2}$}
\begin{picture}(70,20)(0,0)
\thicklines
\put(20,0){\circle{40}}
\put(40,0){\line(1,0){20}}
\put(0,0){\line(-1,-1){15}}
\put(0,0){\line(-1,1){15}}
\end{picture} 
\parbox{1.9cm}{$\Big]_{G_0,\tilde V_0,\tilde U_0}$}
\eea

\vspace*{.5cm}

\xx Substituting (\ref{Vbar1}) and (\ref{V1}) into the terms marked (b) we have:
\bea
\label{(b)}
(b) = -\frac{\hbar^2}{12}\bar V V = -\frac{\hbar^2}{12}V\vG^{-3}V
\eea
The terms marked (c) are explicitly of order $\hbar^3$ which means we can use (\ref{expV}) and (\ref{expU}) to write:
\bea
\label{revsub}
&&(\tilde V_0)_{xyz} = -\bar V_{xyz} \\
&&(\tilde U_0)_{xyzw} = -\bar U_{xyzw} + \bar V_{xya} \vG_{ab} \bar V_{bzw}+ \bar V_{xza} \vG_{ab} \bar V_{byw}+ \bar V_{xwa} \vG_{ab} \bar V_{bzy} \nonumber
\eea
where we have written out the indices explicitly to emphasize the fact that one must be careful to correctly identify the diagrams that are produced when the substitution (\ref{revsub}) is made. 
Substituting we obtain:
\bea
\label{(c)}
\parbox{2.5cm}{$(c) = -i\hbar^3 \Big[ \frac{1}{48} $}
\begin{picture}(75,20)(0,0)
\thicklines
\put(20,0){\circle{40}}
\put(45,0){\circle{40}}
\end{picture}
\parbox{1.2cm}{$+\frac{1}{12}~\Big($}
\begin{picture}(55,20)(0,0)
\thicklines
\put(20,0){\circle{40}}
\put(0,0){\line(1,0){40}}
\put(20,0){\line(0,-1){20}}
\end{picture}
\parbox{1cm}{$-\frac{3}{2}$}
\begin{picture}(55,20)(0,0)
\thicklines
\put(20,0){\circle{40}}
\put(10,0){\circle{20}}
\put(20,0){\line(1,0){20}}
\end{picture}
\parbox{.8cm}{$\Big)\Big]_{\vG,\bar V,\bar U}$}
\eea

\vspace*{5mm}

\xx Combining (\ref{(b)}) and (\ref{(c)}) with the terms marked (e) in (\ref{Gamma-4b}) we have:
\bea
\label{part2}
\Gamma[\vp,\vG,V,U]_{{\rm Part}\,2} = &&-\frac{\hbar^2}{3!}V S_c^{(3)}- i\frac{\hbar^3}{4!}U S_c^{(4)}-\frac{\hbar^2}{8}\vG^2S_c^{(4)} \parbox{1.2cm}{$-\frac{\hbar^2}{12}\Big[$}
\begin{picture}(45,20)(0,0)
\thicklines
\put(20,0){\circle{40}}
\put(0,0){\line(40,0){40}}
\end{picture}
\parbox{2cm}{$\Big]_{\vG,\bar V,\bar U}$}\\[6mm]
&&\parbox{1.5cm}{$-i\hbar^3 \Big[ \frac{1}{48} $}
\begin{picture}(75,20)(0,0)
\thicklines
\put(20,0){\circle{40}}
\put(45,0){\circle{40}}
\end{picture}
\parbox{1.2cm}{$+\frac{1}{12}~\Big($}
\begin{picture}(55,20)(0,0)
\thicklines
\put(20,0){\circle{40}}
\put(0,0){\line(1,0){40}}
\put(20,0){\line(0,-1){20}}
\end{picture}
\parbox{1cm}{$-\frac{3}{2}$}
\begin{picture}(55,20)(0,0)
\thicklines
\put(20,0){\circle{40}}
\put(10,0){\circle{20}}
\put(20,0){\line(1,0){20}}
\end{picture}
\parbox{.8cm}{$\Big)\Big]_{\vG,\bar V,\bar U}$}\nonumber 
\eea

\vspace*{5mm}

\xx The final result for $\Gamma^{(3)}[\vp,\vG,V,U]$ is given by the sum of (\ref{part1}) and (\ref{part2}).

\vspace*{.2cm}

\subsection{Equations of Motion}

\xx To interpret this result we calculate the corresponding equations of motion by taking functional derivatives with respect to $\vG$, $V$ and $U$, and setting the sources to zero. We rewrite the result by using the fact that we have the freedom to take $\bar V \rightarrow \bar V^{(0)}$ and $\bar U \rightarrow \bar U^{(0)}$ anywhere we choose in the terms that carry an explicit factor of $\hbar^3$. Using (\ref{expV}) and (\ref{expU}) we have
\bea
\label{barerevsub}
&&(\bar V_0)_{xyz} = -(\tilde V_0)_{xyz} \\
&&(\bar U_0)_{xyzw} = -(\tilde U_0)_{xyzw} + (\tilde V_0)_{xya} \vG_{ab} (\tilde V_0)_{bzw}+ (\tilde V_0)_{xza} \vG_{ab} (\tilde V_0)_{byw}+ (\tilde V_0)_{xwa} \vG_{ab} (\tilde V_0)_{bzy} \nonumber
\eea
and from (\ref{prop-4}) we have (dropping 1PR terms)
\bea
\tilde V_0\Big|_{J=B=K=H=0} = S^{(3)}\,;~~~\tilde U_0\Big|_{J=B=K=H=0} = S^{(4)}
\eea
We also note that the effective action must be written as a function of the connected vertices $V$ and $U$, and not the truncated vertices $\bar V$ and $\bar U$, when the functional derivatives are taken. These vertices are related through (\ref{trun-vert}).\\

\xx [1] We take the functional derivative with respect to $\vG_{ab}$ and write out explicitly all indices. From $\Gamma[\vp,\vG,V,U]_{{\rm Part}\,1}$ (as given in (\ref{part1})) we have a contribution:
\bea
\label{1}
i\frac{\hbar}{2}\big(S^{(2)}-\vG^{-1}\big)_{ab}
\eea
Next we consider contributions from $\Gamma[\vp,\vG,V,U]_{{\rm Part}\,2}$ as given in (\ref{part2}).
From the third and fifth terms we have:
\bea
\label{35}
-\frac{\hbar^2}{4}\vG_{xy} S_{axyb}^{(4)}-i\frac{\hbar^3}{12}S^{(4)}_{axyz} \vG_{x\bar x}\vG_{y\bar y}\vG_{z\bar z} \bar U_{\bar x\bar y\bar z b}+i\frac{\hbar^3}{4}S^{(3)}_{axy}\vG_{x\alpha}\vG_{y\bar y}S^{(3)}_{\alpha\beta\gamma}\vG_{\beta \bar x}\vG_{\gamma\bar z}\bar U_{\bar x\bar y\bar z b} 
\eea
Note that in the fifth term we have replaced one of the full vertices by the bare vertex $\bar U_0$ and transformed the bare bar-ed vertex to the tilde-ed vertex using (\ref{barerevsub}). 
The fourth term in (\ref{part2}) gives
\bea
\label{4}
\frac{\hbar^2}{4}\bar V_{axy} \vG_{x\bar x}\vG_{y\bar y} \bar V_{\bar x\bar y b}\Big|_{J=B=K=H=0}
\eea 
Contributions from the sixth and seventh terms can be rearranged to produce:
\bea
\label{67}
-\frac{\hbar^2}{4}\bar V^{(1)}_{axy}\vG_{x \bar x}\vG_{y\bar y}\bar V_{\bar x\bar y b}\Big|_{J=B=K=H=0}-i\frac{\hbar^3}{4}S^{(3)}_{axy}\vG_{x\alpha}\vG_{y\bar y}S^{(3)}_{\alpha\beta\gamma}\vG_{\beta \bar x}\vG_{\gamma\bar z}\bar U_{\bar x\bar y\bar z b}
\eea
where we have dropped contributions that correspond to terms that would come from 2PR contributions to the effective action, since these terms have been dropped from the beginning of the calculation. Note that the last term in (\ref{35}) cancels with the last term in (\ref{67}).
We can combine (\ref{4}) and the first two terms in (\ref{67}) to produce:
\bea
\label{467}
\frac{\hbar^2}{4}\big(\bar V-\bar V^{(1)})_{axy}\vG_{x\bar x}\vG_{y\bar y}\bar V_{\bar x\bar y b} 
\eea
We rewrite this result using 
\bea
(\bar V-\bar V^{(1)}) = \bar V^{(0)} = -\tilde V_0\,;~~~\tilde V_0\Big|_{\{J,B,K,H\}= 0}= S^{(3)}
\eea
and obtain
\bea
\label{467p}
\frac{\hbar^2}{4}S^{(3)}_{axy}\vG_{x\bar x}\vG_{y\bar y}\bar V_{\bar x\bar y b}
\eea
Combining (\ref{1}), (\ref{35}) and (\ref{467p}) gives:
\bea
i\big(S^{(2)}-\vG^{-1}\big)_{ab}-\hbar\Big(\frac{1}{2}S_{axy}^{(3)}\vG_{x\bar x}\vG_{y\bar y}\bar V_{\bar x \bar yb}+\frac{1}{2}\vG_{xy} S_{abxy}^{(4)}+\frac{i\hbar}{6}S_{axyw}^{(4)}\vG_{x\bar x}\vG_{y\bar y}\vG_{z\bar z} U_{xywb}\Big) =0
\eea
Extracting the self energy we have,
\bea
\Pi_{ab} = \big(S^{(2)}-\vG^{-1}\big)_{ab} = -i\hbar\Big(\frac{1}{2}S_{axy}^{(3)}\vG_{x\bar x}\vG_{y\bar y}\bar V_{\bar x \bar yb}+\frac{1}{2}\vG_{xy} S_{abxy}^{(4)}+\frac{i\hbar}{6}S_{axyw}^{(4)}\vG_{x\bar x}\vG_{y\bar y}\vG_{z\bar z} U_{xywb}\Big)
\eea
Graphically this result is represented:
\bea
\label{SDeqn}
\parbox{3cm}{$\Pi = -i\hbar\Big[\frac{1}{2}$}
\begin{picture}(70,20)(0,0)
\thicklines
\put(0,0){\line(-1,0){20}}
\put(20,0){\circle{40}}
\put(40,0){\line(1,0){20}}
\put(40,0){\circle*{10}}
\end{picture}
\parbox{2cm}{$+\frac{1}{2}$}
\begin{picture}(52,20)(0,0)
\thicklines
\put(0,0){\circle{40}}
\put(-20,-20){\line(1,0){40}}
\end{picture}
\parbox{1.7cm}{$+\frac{i\hbar}{6}$}
\begin{picture}(80,20)(0,0)
\thicklines
\put(-20,0){\line(1,0){80}}
\put(20,0){\circle{40}}
\put(40,0){\circle*{10}}
\end{picture}
\parbox{1cm}{$\Big]$}
\eea

\vspace*{6mm}
\xx where the lines represent full propagators $\vG$, the un-blobbed vertices are the bare vertices $S^{(3)}$ or $S^{(4)}$, and the blobbed vertices are the full vertices $\bar V$ or $\bar U$. 
We note that (\ref{SDeqn}) is nothing other than the familiar Schwinger-Dyson equation for the self energy for $\phi^4$ theory \cite{IZ}. \\

\xx [2] The calculation of the functional derivative of the effective action with respect to $V$ can be done similarly. We obtain:
\bea
\label{SDVeqn}
\parbox{4.5cm}{$\bar V_{xyz} = -S^{(3)}_{xyz} -i\frac{\hbar}{2}\Big[~~\begin{array}{l}
x \\ ~ \\  y
\end{array}$}
\begin{picture}(70,20)(0,0)
\thicklines
\put(20,0){\circle{40}}
\put(40,0){\line(1,0){20}}
\put(0,0){\line(-1,-1){15}}
\put(0,0){\line(-1,1){15}}
\put(40,0){\circle*{10}}
\end{picture} 
\parbox{2.1cm}{$z ~+~~~\begin{array}{l}
z \\ ~ \\  x
\end{array}$}
\begin{picture}(70,20)(0,0)
\thicklines
\put(20,0){\circle{40}}
\put(40,0){\line(1,0){20}}
\put(0,0){\line(-1,-1){15}}
\put(0,0){\line(-1,1){15}}
\put(40,0){\circle*{10}}
\end{picture} 
\parbox{2.1cm}{$y ~+~~~\begin{array}{l}
y \\ ~ \\  z
\end{array}$}
\begin{picture}(70,20)(0,0)
\thicklines
\put(20,0){\circle{40}}
\put(40,0){\line(1,0){20}}
\put(0,0){\line(-1,-1){15}}
\put(0,0){\line(-1,1){15}}
\put(00,0){\circle*{10}}
\end{picture} 
\parbox{1cm}{$x ~~\Big]$}
\eea
As in the equation above, the lines represent full propagators $\vG$, the un-blobbed vertices are the bare vertices $S^{(3)}$ or $S^{(4)}$, and the blobbed vertices are the full vertices $\bar V$ or $\bar U$. 
This result is the Schwinger-Dyson equation for the truncated 3-point function in $\phi^4$ theory, to 1-loop order. \\

\xx [3] The calculation of the functional derivative of the effective action with respect to $U$ is straightforward. We obtain
\bea
\label{SDUeqn} 
U_{xyzw} = -S^{(4)}_{xyzw}
\eea
which is just the Schwinger-Dyson equation for the truncated 4-point function in $\phi^4$ theory, at the tree level.
 
\section{Discussion and Conclusions}

In this paper we have worked with $\phi^4$ theory and explicitly demonstrated that there is a connection between the hierarchy of Schwinger-Dyson equations and the equations of motion that one obtains from taking functional derivatives with respect to the variational parameters in a 4PI effective theory: the equations of motion obtained from a 4PI effective action reproduce with the Schwinger-Dyson equations at the loop order immediately below which 5-point vertices appear.

This connection is particularly interesting in the context of gauge invarience. Problems with gauge invarience arise when using a truncated $n$PI effective theory or a truncated hierarchy of SD equations. However, since an $n$PI effective theory simultaneously resums different classes of topologies, one expects that increasing $n$ will reduce gauge dependence, at each order in the expansion scheme.  Our result supports this conclusion since it is clear that truncating the SD equations at higher order in the hierarchy will give a higher degree of accuracy with respect to the Ward identities. Work is currently in progress on an inductive proof of this point. 



\begin{thebibliography}{99}

\bibitem{Ward}
J.M. Luttinger and J.C. Ward, Phys. Rev. {\bf 118}, 1417 (1960).
\bibitem{Lee}
T.D. Lee and C.N. Yang, Phys. rev. {\bf 117}, 22 (1961).
\bibitem{Martin}
P. Martin and C. De Dominicis, J. Math. Phys. {\bf 5}, 14 (1964).
\bibitem{Baym}
G. Baym, Phys. Rev. {\bf 127}, 1391 (1962).
\bibitem{CJT}
J.M. Cornwall, R. Jackiw and E. Tomboulis, Phys. Rev. {\bf D 10}, 2428 (1974).
\bibitem{ASmit}
A. Arrizabalaga and J. Smit, Phys. Rev. {\bf D} {\bf 66}, 065014 (2002).
\bibitem{MEC}
M.E. Carrington, G. Kunstatter and H. Zaraket,  hep-ph/0309084.
\bibitem{Mottola} 
E. Mottola, Proceedings of SEWM 2002, World Scientific publishing (2003), hep-ph/0304279
\bibitem{HvH}
H. van Hees and J. Knoll, Phys.Rev. {\bf D 66}, 025028 (2002).
\bibitem{Norton}
R.E. Norton and J.M. Cornwall, Ann. Phys. {\bf 91}, 106 (1975).
\bibitem{IZ} C. Itzykson and J.-B. Zuber, {\it ``Quantum Field Theory,''} McGraw-Hill, New York (1980).
\end{thebibliography}
\end{document}